# Universal dynamics and microwave control of programmable cavity electro-optic frequency combs

Yunxiang Song[1,†], Tianqi Lei[2,†], Yanyun Xue[2], Andrea Cordaro[1], Michael Haas[1], Guanhao Huang[1], Xudong Li[1], Shengyuan Lu[1], Letícia Magalhães[1], Jiayu Yang[1], Matthew Yeh[1], Xinrui Zhu[1], Neil Sinclair[1], Qihuang Gong[2], Yaowen Hu[2,*], Marko Lončar[1,*]

[1] John A. Paulson School of Engineering and Applied Sciences, Harvard University, Cambridge, MA 02138, USA

[2] State Key Laboratory for Mesoscopic Physics and Frontiers Science Center for Nano-optoelectronics, School of Physics, Peking University, Beijing 100871, China

[†] *These authors contributed equally*

[*] *yaowenhu@pku.edu.cn, loncar@g.harvard.edu*

**Electro-optic (EO) frequency combs are foundational for metrology and spectroscopy. Specifically, microresonator-based cavity EO combs are distinguished by efficient sideband generation, precisely controlled by microwave signals, enabling high-performance integrated frequency references and pulse sources. However, the apparent simplicity of these devices, often described by the EO modulation-induced coupling of nearest-neighbor cavity modes, has limited investigations of their fundamental physics, thereby restricting their full potential. Here, we uncover the universal dynamics and complete frequency lattice connectivity underpinning cavity EO microcombs, as well as characterize the full space of nonlinear optical states, controlled by modulation depth and optical detuning, using the thin-film lithium niobate photonic platform. Leveraging this understanding, we design complex long-range couplings between cavity modes to realize programmable spectro-temporal shaping of the generated combs and pulses. We achieve three technological advances, including repetition-rate flexibility, substantial comb bandwidth extension beyond traditional scaling laws, and resonantly-enhanced flat-top spectrum. Our results provide physical insights for synchronously driven cavity-based EO systems, broadly defined, paving the way for electrically controlled and electrically enhanced comb generators for next-generation photonic applications.**

## Introduction

Coherent mixing between light and microwaves, facilitated by the electro-optic (EO) effect in non-centrosymmetric crystals for instance, has driven many scientific advancements and enabled technological innovations. At the center are electro-optic frequency combs (EOFCs) that provide precise optical frequency generation and control[1–5]. EOFCs produce a large number of evenly spaced sidebands centered around an optical carrier, and the spacing between sidebands is determined by the frequency of the applied microwave signal[6–8]. The simplest EOFC consists of an optical phase modulator, though a combination of several amplitude and phase modulators are commonly used as well[9]. To substantially extend the bandwidth of EOFCs, a phase modulator can be inserted inside an optical resonator; if the microwave modulation frequency matches an integer multiple of the resonator's free spectral range (FSR), sideband generation is resonantly enhanced, producing an ultrabroad frequency comb[10–14]. Such resonant EOFCs, often realized with bulk EO phase modulators inside macroscopic Fabry-Perot or fiber-loop resonators, have been applied to broadband frequency metrology[15], ultrafast pulse generation[16], and simulation of topological physics[17–20]. Compared to more widely investigated frequency combs based on passive mode-locked lasers[21,22], Kerr parametric oscillators[23,24], and cavity solitons[25], resonant EOFCs provide a direct link between the optical and electronic domains, while generally featuring greater operational simplicity, reliability, and flexibility. Despite this, the spectro-temporal properties of resonant EOFCs and their dependence on the underlying system parameters are largely unexplored, and have received significantly less attention than other comb sources[26–30]. Yet, a comprehensive investigation of the formation dynamics of resonant EOFCs is crucial, to uncover new physical phenomena and unlock their full potential for novel applications.

Thin-film lithium niobate (TFLN) photonics, combining strong EO interaction, broad microwave bandwidth, and ultra-low optical losses[31,32] in an integrated platform, has significantly advanced traditional EO device performances and led to new operational regimes. For example, TFLN has enabled terabit coherent communication[33], magnetic-free non-reciprocity[34], quantum-classical data links[35], and ultrafast signal processing[36–38], all on a chip. While TFLN cavity-based EOFCs have featured impressive bandwidths, they have relied on the same physics as their bulk counterparts[39–41], namely microwave driven coupling between nearest-neighbor optical resonances. As a result, the comb states are characterized by double-sided exponentially decaying comb line power in the spectral domain and two circulating optical pulses in the temporal domain. Currently, the effects of strong EO coupling and complex long-range connectivity in the frequency domain remain unknown.

Here, we utilize a different approach to cavity EOFC generation and study their rich dynamics. Our EOFCs operate in the wideband and strong EO coupling limits enabled by TFLN. In this regime, comb spectra and pulse patterns are governed by two parameters – the modulation depth and optical detuning – analogous to the pump power- and detuning-controlled nonlinear optical states in dissipative Kerr resonators[25,42]. We observe abrupt transitions from conventional EO comb states to spectrally-modulated states that correspond to highly structured, multi-pulse dynamics. These transitions occur at discrete modulation depths, each characterized by the spontaneous emergence of additional pulse pairs and a resurgence in the pump-to-comb conversion efficiency. Moreover, optical detuning as large as one-half of the cavity FSR is used to deterministically control the comb state pulse number and spatial separation between pulses. These phenomena are not captured by the conventional description: a one-dimensional lattice of modes that include nearest-neighbor coupling, local dissipation, and coherent drive[40,43]. We developed a simplified physical analysis and a general and universal model[44] to support our observations. Leveraging these insights, we demonstrate three practical advances offered by resonant EOFCs by engineering complex long-range couplings up to the seventh harmonic of the FSR. First, we demonstrate repetition-rate flexible, broadband frequency combs from a single device following conventional bandwidth scaling laws. Second, we demonstrate breaking of this comb span scaling that effectively doubles the comb bandwidth and decreases the comb slope by three-fold, compared to the single-frequency-driven case while consuming the same microwave power. Our results show that utilizing multi-tone electronic drives can substantially boost the performance of resonant EOFCs and even overcome limitations imposed by the photonic circuit, such as low quality factors at short wavelengths relevant to biological imaging[45] and quantum applications[46]. Third, we engineer spectral reflections induced by frequency boundaries to demonstrate resonantly-enhanced flat-top frequency combs with less than 6 dB power variation across more than one hundred frequency modes. Notably, switching between all aforementioned operational modes is possible because the comb state is primarily set by the choice of microwave modulation. Our presented findings yield intuitive physical understanding of resonant EOFCs and provide a rich toolbox for studying cavity electro-optic systems broadly defined. From a practical standpoint, the techniques developed underscore the immense opportunity for resonant EOFCs to provide versatile comb and pulse sources unlimited by the optical carrier frequency, controlled and enhanced by microwave electronics.

## Results

### *Spectro-temporal versatility of resonant electro-optic frequency combs*

Resonant EOFCs in this work are based on traveling-wave modulated racetrack resonators, fabricated on X-cut TFLN-on-insulator wafers. Traveling-wave EO phase modulation facilitated by cm-long coplanar waveguides simultaneously provides strong EO coupling and broad modulation bandwidth. Such properties allow for intense modulation at fast speeds, while consuming moderate microwave power, amounting to strong coupling between cavity modes far beyond nearest-neighbor interactions. Therefore, our platform utilizes a range of microwave signals to uncover unconventional nonlinear optical states and demonstrate active comb state control (Fig. 1).

A continuous-wave optical field ($\omega_p$) excites an optical cavity, consisting of a number of frequency modes ($\omega_\mu$, where subscript $\mu$ labels the mode number). In Fig. 1a, we first consider the conventional case of harmonic phase modulation by a single microwave tone ($\omega_{MW}$), commensurate with the fundamental cavity FSR ($1 \cdot \omega_{FSR}$). When a microwave with voltage amplitude $V_1$ is applied, we define a modulation depth $\beta_1 = \pi \frac{V_1}{V_{\pi,1}}$, where $V_{\pi,1}$ is the voltage amplitude required to impart $\pi$-phase shifts onto intracavity light, at $1 \cdot \omega_{FSR}$ modulation frequency. In the case of small $\beta_1$ and in the absence of optical detuning, nearest-neighbor coupling between resonance modes dominates the comb formation dynamics. The generated comb is characterized by exponential decay of the comb line power and two optical pulses separated by one-half the cavity round trip time ($\frac{1}{2} T_R$, where $T_R = \frac{2\pi}{\omega_{FSR}}$). When $\beta_1$ is increased above $2\pi$, the dynamics become more complex (Fig. 1b). Instead, a strong, single-tone modulation creates many coupling pathways between frequency modes, each carrying distinct amplitudes and phases. In the steady-state, all pathways coherently interfere and lead to comb envelopes exhibiting rapid, periodic oscillations. This substructure indicates additional time-domain periodicity emerging on $T_R$ time scales, e.g., the temporal waveform can show a coherent six-pulse state. Beyond single-tone modulation, multi-tone modulations facilitate programmable dynamics in the frequency domain. Fully accounting for the large set of mutually interfering coupling pathways created by $n$ modulation tones (modulation depths and coupling phases $\{\beta_i, \phi_i\}_{i\in\{1,\ldots,n\}}$), the steady-state output can be designed to satisfy numerous target spectra, including comb bandwidth boosting and resonantly-enhanced flat-top envelopes (Fig. 1c).

### *Observation of unconventional comb and pulse states*

To reveal the resonant EOFC state space, we performed numerical simulations with an extended theoretical model (see Supplementary information for details) and obtained the comb spectrogram as a function of $\beta_1$ (Fig. 2a). In short, the model fully captures the dynamics giving rise to Fig. 1b. As expected, the $0 \le \beta_1 < 2\pi$ regime is characterized by comb states featuring conventional, double-sided exponential decay. The spectrum undergoes transition once $\beta_1$ exceeds $2\pi$, gaining periodic oscillations in its envelope. Such features initially emerge around the pump mode and rapidly disperse into large absolute mode numbers as $\beta_1$ is increased. Similarly, a qualitative change in spectral features and oscillation pattern occurs again at $\beta_1 \sim 4\pi$.

To experimentally observe such comb states and transitions, we characterized a resonant EOFC device with 2-cm EO modulation length, whose $V_{\pi,1}$ and $\omega_{FSR}$ are 5.5 V and $2\pi \cdot 3.141$ GHz, respectively. Applying a single-tone microwave with $\omega_{MW} = 1 \cdot \omega_{FSR}$, we increase the microwave power such that $\beta_1$ is swept from $0.096\pi$ up to $2.863\pi$ (Fig. 2b). This estimated range is indicated by the dashed white box in the Fig. 2a simulation, and it is determined by the EO response calibration of the full modulation circuit. The simulated and measured spectrograms show excellent agreement. The simulation also predicts that the pump-to-comb conversion efficiency ($\eta$) undergoes dramatic revivals at comb state transition boundaries, instead of monotonically decreasing with $\beta_1$ as previously observed[41]. Such revivals are confirmed in experiment by summing the peak values of all EOFC sidebands (Fig. 2c) relative to the out-coupled pump power.

Revivals in $\eta$ can be readily explained by analyzing the temporal dynamics of resonant EOFCs as a function of $\beta_1$ (Fig. 2d). We note that the critical value of $\beta_1 \sim 2\pi$ is associated with the spontaneous emergence of additional pulse pairs. As $\beta_1$ is increased from $2\pi$ to $4\pi$, each additional pulse pair converges towards the original two pulses in the $\beta_1 < 2\pi$ regime, while all pulses continuously experience narrowing in temporal width. Reconsidering the corresponding spectra, the periodic substructure in spectral envelopes exhibits a smoothly increasing periodicity with $\beta_1$, consistent with the smooth temporal (spatial) reorganization of six pulses within the cavity. Much like multi-soliton states in dissipative Kerr resonators[42], the $\eta$ is higher in the many-pulse regime for resonant EOFCs as well. For clarity, we isolate three slices of the experimental spectrogram (Fig. 2e) to showcase representative comb state evolution from conventional EOFCs into coherent six-pulse states, achieved by simply increasing $\beta_1$.

### *Observation of highly-detuned comb and pulse states*

Next, we explore the impact of optical detuning ($\Delta_p = \omega_p - \omega_{\mu=0}$) on the EOFC state space. A key requirement for associating measured comb states with particular $\Delta_p$s is the ability to probe $\Delta_p$ during comb generation. However, steady-state transmission measurements of a strongly modulated EO comb source contain no information of the pumped cavity modes, in contrast to traditional detuning spectroscopies. In the Supplementary information, we introduce a method to determine $\Delta_p$ from the frequency degenerate counter-clockwise (CCW) modes, exploiting the traveling-wave isolation effect[47,48]. We find that $\Delta_p$ can be accurately mapped by constantly observing the CCW band-structure, even up to the highest $\beta_1$ attempted. Utilizing this method, we numerically and experimentally study comb state evolution with $\beta_1$ when optical detuning is maximized ($\omega_p$ centered between e.g., $\omega_0$ and $\omega_1$). Figure 2f shows the simulated comb spectrogram as a function of $\beta_1$, when $\Delta_p = \frac{\omega_{FSR}}{2}$. Comb formation occurs immediately once $\beta_1$ exceeds the threshold of $\pi$, and the emerging comb spectrum is characterized by modulated envelopes in contrast to pure exponential decay. Comb state transitions are also predicted to occur at $\beta_1 \sim 3\pi$ and $5\pi$. Using the previous device, we again apply a single-tone microwave with $\omega_{MW} = 1 \cdot \omega_{FSR}$ and measure the spectrogram for calibrated $\beta_1$s between $0.356\pi$ to $3.052\pi$ (Fig. 2g). The temporal dynamics in this case indicates the spontaneous emergence of four pulses instead of two (Fig. 2i), originating from an initially empty cavity. Subsequently, four-pulse (eight-pulse) states are created in the $\pi \le \beta_1 < 3\pi$ ($3\pi \le \beta_1 < 5\pi$) regime, as shown in Fig. 2j. We note that the study of eight-pulse states was limited by the available microwave power.

The essential intuition for comb existence under extreme optical detunings involves the expansion and contraction of the optical path length of the cavity, understood as collective resonance oscillations given by amplitude $\Delta\omega_\mu = \frac{\beta_1}{2\pi} \cdot \omega_{FSR}$. Thus, each oscillation covers a frequency band of approximately $2\Delta\omega_\mu$. When $2\Delta\omega_\mu < \omega_{FSR}$ (i.e., $\beta_1 < \pi$), these bands are mutually isolated, and band gaps are formed. No comb can be generated when $\omega_p$ lies within these gaps, hence the resonant EOFC has finite existence range. On the other hand, when $2\Delta\omega_\mu > \omega_{FSR}$ ($\beta_1 > \pi$), these bands overlap and $\omega_p$ will always couple into the cavity, enabling detuning agnostic operation. When $\Delta_p = \frac{\omega_{FSR}}{2}$, the initial EOFC state at $\beta_1 = \pi$ arises from two resonances instead of one (each contributing two pulses), owing to the symmetry of the optical detuning with respect to oscillating resonances. The same intuition explains the transitions from conventional EOFC states into larger pulse numbers and coherent temporal structures, when $\Delta_p = 0$. For the harmonically-driven cavity EO system, it is important to note that resonances are well-defined in the unitless slow-time ($\tau = t/T_R$). Surprisingly, this time-domain picture accurately predicts EO comb spans and pulse widths from simple scaling, conceptualizing them as solutions to a classical mechanics problem (see Supplementary information for detailed discussion). This connection strongly informs multi-tone electronic drive design and analysis, which we describe next.

### ***Repetition-rate flexible and broadband combs***

We next leverage complex long-range coupling dynamics in the frequency domain to demonstrate advanced functionalities. The simplest form of long-range coupling can be achieved using a single-tone drive with frequency corresponding to a multiple of the FSR, giving rise to frequency combs with tunable spectral spacing and span, of interest for many applications[49]. Importantly, broad EO bandwidth allows for efficient multi-FSR modulations of the cavity, and repetition-rate flexible resonant EOFCs from a single device are possible. We modified our resonant EOFC device to include a dispersion engineered waveguide featuring equally spaced cavity modes over a large optical bandwidth and reduced the modulator length to 1.45 cm to achieve relatively flat EO response. We measured $V_{\pi,1}$ and $\omega_{FSR}$ to be 7.12 V and $2\pi \cdot 4.187$ GHz, respectively, as well as $V_{\pi,n}$ at $n^{\text{th}}$ harmonics of the fundamental FSR up to $n = 10$ (Fig. 3a). The 3 dB EO response is inferred to be $\omega_{MW,3dB} \sim 2\pi \cdot 33.5$ GHz (Fig. 3b). Selecting $n = 4, 5,$ and 7, we generate combs with about 1 mW of optical power on-chip, modulation frequencies $\omega_{MW} = n \cdot \omega_{FSR} = 2\pi \cdot 16.750, 20.935,$ and $29.312$ GHz (Figs. 3c, d), while maintaining approximately constant $\beta_n \sim \pi$ (calibration uses a fiber Bragg grating to observe existence range merging). Our device has a finesse of $F \sim 42.65$, and we observe that the number of comb modes is constant (~665). This is consistent with our derived scaling law of $-\left(\frac{\beta_n F}{2\pi}\right)^{-1}$ in the $0 \leq \beta_n < 2\pi$ regime. Thus, the total span scales linearly with the comb spacing in frequency units, given by 11.26, 13.90, 19.76 THz spans around a center frequency of $f_p \sim 194.67$ THz (equivalently, 90, 109, and 156 nm spans around a center wavelength of $\lambda_p \sim 1540$ nm). Such scaling indicates that a flat EO response would amount to identical comb generation dynamics for all modulation harmonics and can be treated equally.

### *Multi-tone comb boosting and resonantly-enhanced flat-top combs*

Multi-tone drives, with actively tunable modulation depths and phases, can result in even richer complex long-range couplings between frequency modes of the cavity (Fig. 4a). This network of higher-order connectivity inherently influences the resonant EO comb and pulse state, which can offer several practical advantages. One example is flattening of the generated spectrum and narrowing of the resulting optical pulse, resulting in substantial extension of comb bandwidth compared to what is achievable using a single-tone drive. Towards this goal, we synthesize a four-tone complex waveform $\sum_{n=1}^{7} V_n' \cos(n \cdot \omega_{FSR} t + \phi_n')$, where $V_1', V_3', V_5',$ and $V_7'$ are $6.46, 2.15, 1.29,$ and $0.92$ V, respectively, $\phi_n'$ are all $\frac{\pi}{2}$, and $V_2', V_4', V_6'$ are all 0 V. This amounts to a square wave with fundamental frequency $1 \cdot \omega_{FSR}$, subject to a brick-wall filter at $\omega_{MW,3dB}$. The frequency content of this waveform covers seven harmonics of the cavity FSR. We denote the total microwave power by $\sum_{n=1}^{7} P_n'$, where $P_n'$ is proportional to $|V_n'|^2$. On the other hand, a single-tone modulation with $P_1 = \sum_{n=1}^{7} P_n'$ corresponds to a $V_1$ of 6.99 V. Remarkably, as shown in Fig. 4b, when separately applying these modulation waveforms to our device, the multi-tone modulation results in 457 modes and 2.9 ps pulse width (including fiber dispersion in the detection path), which is approximately double the comb span and half the pulse width generated by a single-tone modulation (216 modes and 5.7 ps). This is a physics-informed choice obtained from the time-domain picture, where EOFC pulse width (and hence comb span) is directly mapped to electronic rise times. In the Supplementary information, we derive much improved comb span and pulse width scaling laws, achievable in higher-order modulation scenarios. We note that in this experiment, we modified the device from the previous sections by incorporating an auxiliary ring and a drop port waveguide, which suppresses the background pump and allows for pulse measurements via intensity autocorrelation.

Given our finding, we further explore the potential for resonantly-enhanced flat-top EOFCs. In conjunction with the multi-tone modulation as described, frequency domain boundaries can further enhance comb flatness. Intuitively, the modulation efficiently injects light away from the pump. A boundary is reached when sidebands cascade far enough, such that their frequencies no longer satisfy the resonance condition of the cavity. In other words, optical photons that encounter this boundary can only propagate back towards the pump and undergo coherent interference with existing modes closer to the pump, thereby flattening the comb. Here, we create such boundaries using microwave detuning, equivalent to a synthetic linear dispersion. We modify the complex long-range couplings such that $\sum_{n=1}^{7} V_n' \cos\left(n \cdot (\omega_{FSR} + \Delta_{MW})t + \phi_n'\right)$ features nonzero $\Delta_{MW}$. Specifically, we vary $\Delta_{MW}$ between 10.1 and 18.3 MHz, noting flatness enhancement, as well as sharp roll offs in comb power indicative of boundary reflections taking effect (Fig. 4c). We then vary the microwave power while keeping the boundary fixed ($\Delta_{MW} = 14.2$ MHz) and the relative voltage ratios constant. While the total number of modes steadily increases, the comb slope saturates to approximately 0.035 dB/mode at around 29.1 dBm of applied microwave power (Fig. 4d). It is worth nothing that with a single-tone modulation, to achieve identical levels of spectral flatness on the same device would require a loaded quality factor exceeding 9 million, provided with the same microwave power. Remarkably, the device used in this measurement has a loaded quality factor of only 1.14 million, highlighting that long-range coupling design and synthetic frequency dimension concepts can be readily applied to relax stringent requirements on the optical circuit quality, to achieve certain comb metrics.

## Discussion

In summary, we conducted a comprehensive study of the state space of resonant EOFCs leveraging the TFLN platform, uncovering transitions from conventional comb states to those characterized by highly-modulated spectral envelopes and temporally-structured pulse patterns. We identified modulation depth and optical detuning as key order parameters in the state space, and we developed a pump-probe spectroscopic method to systematically measure the latter during comb operation. Our findings have implications on both resonant EOFC limits and opportunities. For example, in the case that minimal comb line power variation over large optical bandwidth is desired, the microwave power in a single-tone cannot be arbitrarily increased to broaden comb span and only improvements in optical loss is effective. For much broader combs than those currently achieved, Kerr[50–55] or hybrid-nonlinear[56,57] approaches should be considered. The latter is synergistic with our findings here, as resonant EOFCs operating in the detuning-agnostic regime can efficiently generate sidebands from widely spaced, multi-wavelength sources, such as THz-spaced Kerr solitons, without any need for precise frequency matching between light source and cavity modes. On the other hand, higher-order comb and pulse states featuring greater conversion efficiencies may be exploited for broadband applications where line-by-line power variation is not essential, such as in optical frequency metrology, or in ultrafast processes requiring effectively high repetition-rate, pulsed excitations. The physical insight and the modeling framework we established[58] may also guide the exploration of new frontiers in cavity EO devices operating in the strong and broadband EO coupling regime, including novel frequency conversion devices[59,60], Pockels-enhanced light sources[61,62], microwave-to-optical transducers[63], and actively mode-locked lasers[64].

The utilization of complex long-range couplings in the context of resonant EOFCs led to repetition-rate flexible comb sources, as well as comb bandwidth boosting and resonantly-enhanced flat-top comb generation methods. Active control of the modulation frequencies and total waveform thus endows an exceptional reconfigurability and programmability to the resonant EOFC generation modality. This stands in contrast with meta-dispersion-shaped Kerr combs[65,66] which are not only fixed after fabrication but also stochastic in terms of the states supported by any given device and its resonances. In future developments, decreasing the cavity's fundamental FSR could provide finer control over the comb's repetition-rate, while increasing the modulation bandwidth could facilitate even larger scale complex connectivity in the frequency domain. From a fundamental science perspective, quantum simulations of time-harmonic systems[40,67] and non-Hermitian physics[17,18,68] could benefit from the extensive connectivity enabled by this platform. On a practical level, the vast design space created by the combination of distinct coupling amplitudes and phases can be explored using inverse design and computational optimization[69]. These methods could efficiently optimize free parameters to achieve a desired comb profile, towards highly versatile and multi-functional devices. As a first demonstration, our physics-informed optimization already produces combs substantially surpassing the conventional comb span scaling laws, otherwise limited by cavity finesse and microwave power. Integrated comb sources featuring many thousands of modes separated by a GHz-level spacing should be well within reach, by simply doubling of the loaded quality factor of our present device – state-of-the-art TFLN racetracks resonators can improve optical losses by about ten-fold[70]. Our results highlight TFLN cavity EOFCs as a unique candidate to fill in the low repetition-rate metric space of integrated frequency combs, hard to achieve through other nonlinearities[71]. They also point towards the development of high-performance combs across a range of pump wavelengths, leveraging microwave electronic enhancement via multi-harmonic driving, particularly at short wavelengths where parametric methods face significant challenges imposed by high optical losses, inadequate dispersion conditions, and lack of amplification techniques. Through active microwave control, the resonant EOFCs can be adapted on-demand to fulfill various applications, including comb-based spectroscopy[72–75], arbitrary waveform generation[76,77], and optical communication[78–81].

## References

1. Li, J., Yi, X., Lee, H., Diddams, S. A. & Vahala, K. J. Electro-optical frequency division and stable microwave synthesis. *Science* **345**, 309–313 (2014).
2. Karpiński, M., Jachura, M., Wright, L. J. & Smith, B. J. Bandwidth manipulation of quantum light by an electro-optic time lens. *Nat. Photonics* **11**, 53–57 (2017).
3. Beha, K. *et al.* Electronic synthesis of light. *Optica* **4**, 406 (2017).
4. Carlson, D. R. *et al.* Ultrafast electro-optic light with subcycle control. *Science* **361**, 1358–1363 (2018).
5. Moille, G. *et al.* Kerr-induced synchronization of a cavity soliton to an optical reference. *Nature* **624**, 267–274 (2023).
6. Torres-Company, V. & Weiner, A. M. Optical frequency comb technology for ultra-broadband radio-frequency photonics. *Laser Photonics Rev.* **8**, 368–393 (2014).
7. Parriaux, A., Hammani, K. & Millot, G. Electro-optic frequency combs. *Adv. Opt. Photonics* **12**, 223 (2020).
8. Chang, L., Liu, S. & Bowers, J. E. Integrated optical frequency comb technologies. *Nat. Photonics* **16**, 95–108 (2022).
9. Metcalf, A. J., Torres-Company, V., Leaird, D. E. & Weiner, A. M. High-Power Broadly Tunable Electrooptic Frequency Comb Generator. *IEEE J. Sel. Top. Quantum Electron.* **19**, 231–236 (2013).
10. Kobayashi, T., Sueta, T., Cho, Y. & Matsuo, Y. High-repetition-rate optical pulse generator using a Fabry-Perot electro-optic modulator. *Appl. Phys. Lett.* **21**, 341–343 (1972).
11. Ho, K.-P. & Kahn, J. M. Optical frequency comb generator using phase modulation in amplified circulating loop. *IEEE Photonics Technol. Lett.* **5**, 721–725 (1993).

12. Kourogi, M., Enami, T. & Ohtsu, M. A monolithic optical frequency comb generator. *IEEE Photonics Technol. Lett.* **6**, 214–217 (1994).

13. Saitoh, T. *et al.* Modulation characteristic of waveguide-type optical frequency comb generator. *J. Light. Technol.* **16**, 824–832 (1998).

14. Buscaino, B., Zhang, M., Loncar, M. & Kahn, J. M. Design of Efficient Resonator-Enhanced Electro-Optic Frequency Comb Generators. *J. Light. Technol.* **38**, 1400–1413 (2020).

15. Kourogi, M., Nakagawa, K. & Ohtsu, M. Wide-span optical frequency comb generator for accurate optical frequency difference measurement. *IEEE J. Quantum Electron.* **29**, 2693–2701 (1993).

16. Xiao, S., Hollberg, L., Newbury, N. R. & Diddams, S. A. Toward a low-jitter 10 GHz pulsed source with an optical frequency comb generator. *Opt. Express* **16**, 8498 (2008).

17. Wang, K. *et al.* Generating arbitrary topological windings of a non-Hermitian band. *Science* **371**, 1240–1245 (2021).

18. Wang, K., Dutt, A., Wojcik, C. C. & Fan, S. Topological complex-energy braiding of non-Hermitian bands. *Nature* **598**, 59–64 (2021).

19. Englebert, N. *et al.* Bloch oscillations of coherently driven dissipative solitons in a synthetic dimension. *Nat. Phys.* **19**, 1014–1021 (2023).

20. Kim, S., Krasnok, A. & Alù, A. Complex-frequency excitations in photonics and wave physics. *Science* **387**, eado4128 (2025).

21. Udem, Th., Reichert, J., Holzwarth, R. & Hänsch, T. W. Absolute Optical Frequency Measurement of the Cesium D 1 Line with a Mode-Locked Laser. *Phys. Rev. Lett.* **82**, 3568–3571 (1999).

22. Jones, D. J. *et al.* Carrier-Envelope Phase Control of Femtosecond Mode-Locked Lasers and Direct Optical Frequency Synthesis. *Science* **288**, 635–639 (2000).

23. Del'Haye, P. *et al.* Optical frequency comb generation from a monolithic microresonator. *Nature* **450**, 1214–1217 (2007).

24. Okawachi, Y. *et al.* Octave-spanning frequency comb generation in a silicon nitride chip. *Opt. Lett.* **36**, 3398 (2011).

25. Herr, T. *et al.* Temporal solitons in optical microresonators. *Nat. Photonics* **8**, 145–152 (2014).

26. Herr, T. *et al.* Universal formation dynamics and noise of Kerr-frequency combs in microresonators. *Nat. Photonics* **6**, 480–487 (2012).

27. Coen, S., Randle, H. G., Sylvestre, T. & Erkintalo, M. Modeling of octave-spanning Kerr frequency combs using a generalized mean-field Lugiato–Lefever model. *Opt. Lett.* **38**, 37 (2013).

28. Lamont, M. R. E., Okawachi, Y. & Gaeta, A. L. Route to stabilized ultrabroadband microresonator-based frequency combs. *Opt. Lett.* **38**, 3478 (2013).

29. Chembo, Y. K. & Menyuk, C. R. Spatiotemporal Lugiato-Lefever formalism for Kerr-comb generation in whispering-gallery-mode resonators. *Phys. Rev. A* **87**, 053852 (2013).

30. Guo, H. *et al.* Universal dynamics and deterministic switching of dissipative Kerr solitons in optical microresonators. *Nat. Phys.* **13**, 94–102 (2017).

31. Zhu, D. *et al.* Integrated photonics on thin-film lithium niobate. *Adv. Opt. Photonics* **13**, 242 (2021).

32. Boes, A. *et al.* Lithium niobate photonics: Unlocking the electromagnetic spectrum. *Science* **379**, eabj4396 (2023).

33. Xu, M. *et al.* Dual-polarization thin-film lithium niobate in-phase quadrature modulators for terabit-per-second transmission. *Optica* **9**, 61 (2022).

34. Herrmann, J. F. *et al.* Mirror symmetric on-chip frequency circulation of light. *Nat. Photonics* **16**, 603–608 (2022).

35. Shen, M. *et al.* Photonic link from single-flux-quantum circuits to room temperature. *Nat. Photonics* **18**, 371–378 (2024).

36. Feng, H. *et al.* Integrated lithium niobate microwave photonic processing engine. *Nature* **627**, 80–87 (2024).

37. Hu, Y. *et al.* Integrated lithium niobate photonic computing circuit based on efficient and high-speed electro-optic conversion. Preprint at https://doi.org/10.48550/arXiv.2411.02734 (2024).

38. Song, Y. *et al.* Integrated electro-optic digital-to-analog link for efficient computing and arbitrary waveform generation. Preprint at https://doi.org/10.48550/arXiv.2411.04395 (2024).

39. Zhang, M. *et al.* Broadband electro-optic frequency comb generation in a lithium niobate microring resonator. *Nature* **568**, 373–377 (2019).

40. Hu, Y., Reimer, C., Shams-Ansari, A., Zhang, M. & Loncar, M. Realization of high-dimensional frequency crystals in electro-optic microcombs. *Optica* **7**, 1189 (2020).

41. Hu, Y. *et al.* High-efficiency and broadband on-chip electro-optic frequency comb generators. *Nat. Photonics* **16**, 679–685 (2022).

42. Godey, C., Balakireva, I. V., Coillet, A. & Chembo, Y. K. Stability analysis of the spatiotemporal Lugiato-Lefever model for Kerr optical frequency combs in the anomalous and normal dispersion regimes. *Phys. Rev. A* **89**, 063814 (2014).

43. Stokowski, H. S. *et al.* Integrated frequency-modulated optical parametric oscillator. *Nature* **627**, 95–100 (2024).

44. Lei, T. *et al.* Strong-coupling and high-bandwidth cavity electro-optic modulation for advanced pulse-comb synthesis. Preprint at https://arxiv.org/abs/2507.21855 (2025).

45. Karpov, M., Pfeiffer, M. H. P., Liu, J., Lukashchuk, A. & Kippenberg, T. J. Photonic chip-based soliton frequency combs covering the biological imaging window. *Nat. Commun.* **9**, 1146 (2018).

46. Isichenko, A. *et al.* Photonic integrated beam delivery for a rubidium 3D magneto-optical trap. *Nat. Commun.* **14**, 3080 (2023).

47. Sounas, D. L. & Alù, A. Non-reciprocal photonics based on time modulation. *Nat. Photonics* **11**, 774–783 (2017).

48. Yu, M. *et al.* Integrated electro-optic isolator on thin-film lithium niobate. *Nat. Photonics* **17**, 666–671 (2023).

49. Millot, G. *et al.* Frequency-agile dual-comb spectroscopy. *Nat. Photonics* **10**, 27–30 (2016).

50. Li, Q. *et al.* Stably accessing octave-spanning microresonator frequency combs in the soliton regime. *Optica* **4**, 193 (2017).

51. Pfeiffer, M. H. P. *et al.* Octave-spanning dissipative Kerr soliton frequency combs in Si_3N_4 microresonators. *Optica* **4**, 684 (2017).

52. Liu, X. *et al.* Aluminum nitride nanophotonics for beyond-octave soliton microcomb generation and self-referencing. *Nat. Commun.* **12**, 5428 (2021).

53. Weng, H. *et al.* Directly accessing octave-spanning dissipative Kerr soliton frequency combs in an AlN microresonator. *Photonics Res.* **9**, 1351 (2021).

54. Wang, P.-Y. *et al.* Octave soliton microcombs in lithium niobate microresonators. *Opt. Lett.* **49**, 1729 (2024).

55. Song, Y., Hu, Y., Zhu, X., Yang, K. & Lončar, M. Octave-spanning Kerr soliton frequency combs in dispersion- and dissipation-engineered lithium niobate microresonators. *Light Sci. Appl.* **13**, 225 (2024).

56. Drake, T. E. *et al.* Terahertz-Rate Kerr-Microresonator Optical Clockwork. *Phys. Rev. X* **9**, 031023 (2019).

57. Song, Y., Hu, Y., Lončar, M. & Yang, K. Hybrid Kerr-electro-optic frequency combs on thin-film lithium niobate. Preprint at http://arxiv.org/abs/2402.11669 (2024).

58. Lei, T. *et al.* Strong-coupling and high-bandwidth cavity electro-optic modulation for machine-learning-enabled pulse-comb synthesis. *Submitted*.

59. Zhang, M. *et al.* Electronically programmable photonic molecule. *Nat. Photonics* **13**, 36–40 (2019).

60. Hu, Y. *et al.* On-chip electro-optic frequency shifters and beam splitters. *Nature* **599**, 587–593 (2021).

61. Li, M. *et al.* Integrated Pockels laser. *Nat. Commun.* **13**, 5344 (2022).

62. Ling, J. *et al.* Electrically empowered microcomb laser. *Nat. Commun.* **15**, 4192 (2024).

63. Holzgrafe, J. *et al.* Cavity electro-optics in thin-film lithium niobate for efficient microwave-to-optical transduction. *Optica* **7**, 1714 (2020).

64. Guo, Q. *et al.* Ultrafast mode-locked laser in nanophotonic lithium niobate. *Science* **382**, 708–713 (2023).

65. Lucas, E., Yu, S.-P., Briles, T. C., Carlson, D. R. & Papp, S. B. Tailoring microcombs with inverse-designed, meta-dispersion microresonators. *Nat. Photonics* **17**, 943–950 (2023).

66. Moille, G., Lu, X., Stone, J., Westly, D. & Srinivasan, K. Fourier synthesis dispersion engineering of photonic crystal microrings for broadband frequency combs. *Commun. Phys.* **6**, 144 (2023).

67. Javid, U. A. *et al.* Chip-scale simulations in a quantum-correlated synthetic space. *Nat. Photonics* **17**, 883–890 (2023).

68. Feng, L., El-Ganainy, R. & Ge, L. Non-Hermitian photonics based on parity–time symmetry. *Nat. Photonics* **11**, 752–762 (2017).

69. Molesky, S. *et al.* Inverse design in nanophotonics. *Nat. Photonics* **12**, 659–670 (2018).

70. Zhu, X. *et al.* Twenty-nine million intrinsic $Q$ -factor monolithic microresonators on thin-film lithium niobate. *Photonics Res.* **12**, A63 (2024).

71. Suh, M.-G. & Vahala, K. Gigahertz-repetition-rate soliton microcombs. *Optica* **5**, 65 (2018).

72. Suh, M.-G., Yang, Q.-F., Yang, K. Y., Yi, X. & Vahala, K. J. Microresonator soliton dual-comb spectroscopy. *Science* **354**, 600–603 (2016).

73. Yi, X., Yang, Q.-F., Yang, K. Y. & Vahala, K. Imaging soliton dynamics in optical microcavities. *Nat. Commun.* **9**, 3565 (2018).

74. Shams-Ansari, A. *et al.* Thin-film lithium-niobate electro-optic platform for spectrally tailored dual-comb spectroscopy. *Commun. Phys.* **5**, 88 (2022).

75. Long, D. A., Stone, J. R., Sun, Y., Westly, D. & Srinivasan, K. Sub-Doppler spectroscopy of quantum systems through nanophotonic spectral translation of electro-optic light. *Nat. Photonics* (2024) doi:10.1038/s41566-024-01532-w.

76. Cundiff, S. T. & Weiner, A. M. Optical arbitrary waveform generation. *Nat. Photonics* **4**, 760–766 (2010).

77. Wang, B., Yang, Z., Sun, S. & Yi, X. Radio-frequency line-by-line Fourier synthesis based on optical soliton microcombs. *Photonics Res.* **10**, 932 (2022).

78. Marin-Palomo, P. *et al.* Microresonator-based solitons for massively parallel coherent optical communications. *Nature* **546**, 274–279 (2017).

79. Fülöp, A. *et al.* High-order coherent communications using mode-locked dark-pulse Kerr combs from microresonators. *Nat. Commun.* **9**, 1598 (2018).

80. Jørgensen, A. A. *et al.* Petabit-per-second data transmission using a chip-scale microcomb ring resonator source. *Nat. Photonics* **16**, 798–802 (2022).

81. Rizzo, A. *et al.* Massively scalable Kerr comb-driven silicon photonic link. *Nat. Photonics* **17**, 781–790 (2023).

**Acknowledgements:** Y.S. thanks Nicholas Achuthan, Rebecca Cheng, Tong Ge, Brandon Grinkemeyer, Jingchao Fang, Prof. Evelyn Hu, Keith Powell, Shima Rajabali, Dylan Renaud, Nicholas Rivera, Prof. Thomas Rosenbaum, Urban Senica, Amirhassan Shams-Ansari, Anna Shelton, Jamison Sloan, Hana Warner, Chenjie Xin, Zhiquan Yuan, and Xiangying Zuo for discussions. Y.S. thanks Ganna Savostyanova and Meg Reardon for administrative support, as well as Tom Chen (Flexcompute), Kyle Richard, Neil Hoffman, and Mark Roberts (Keysight Technologies, Inc.) for technical support. The device fabrication in this work was performed at the Harvard University Center for Nanoscale Systems (CNS); a member of the National Nanotechnology Coordinated Infrastructure Network (NNCI), which is supported by the National Science Foundation under NSF award no. ECCS-2025158.

**Author contributions:** Y.H., M.L., Y.S., T.L. conceived the project. Y.S. designed the devices with help from Y.H. Y.S. performed device fabrication and experimental characterizations. Y.S. and T.L. performed the numerical simulations with help from Y.X. Y.S., T.L., Y.X., Y.H., and M.L. interpreted and analyzed the data. A.C., M.H., G.H., X.L., S.L., L.M., N.S., J.Y., M.Y., and X.Z. provided helpful assistance on all aspects of the project. Y.S. and M.L. wrote the manuscript with contributions from all authors. M.L., Y.H., and Q.G. supervised the project.

**Funding:** This work was supported by National Research Foundation funded by the Korea government (NRF-2022M3K4A1094782); AFOSR (FA955024PB004); National Science Foundation (OMA-2138068; OMA-2137723; EEC-1941583); NASA (80NSSC22K0262); Department of the Navy (N6833522C0413); Amazon Web Services (A50791). A.C. acknowledges Rubicon postdoctoral fellowship from the Netherlands Organization for Scientific Research (019.231EN.011). G.H. acknowledges financial support from the Swiss National Science Foundation (Postdoc. Mobility, 222257). L.M. acknowledges funding from Behring foundation and CAPES-Fulbright. S.L. acknowledges doctoral fellowship from A-STAR.

**Competing interests:** M.L. is involved in developing lithium niobate technologies at HyperLight Corporation. The authors declare no other competing interests.

**Data and materials availability:** All data needed to evaluate the conclusions in the paper are present in the paper and/or the Supplementary Information.

**List of supplementary materials:**

Supplementary Text

Figs. S1 to S13

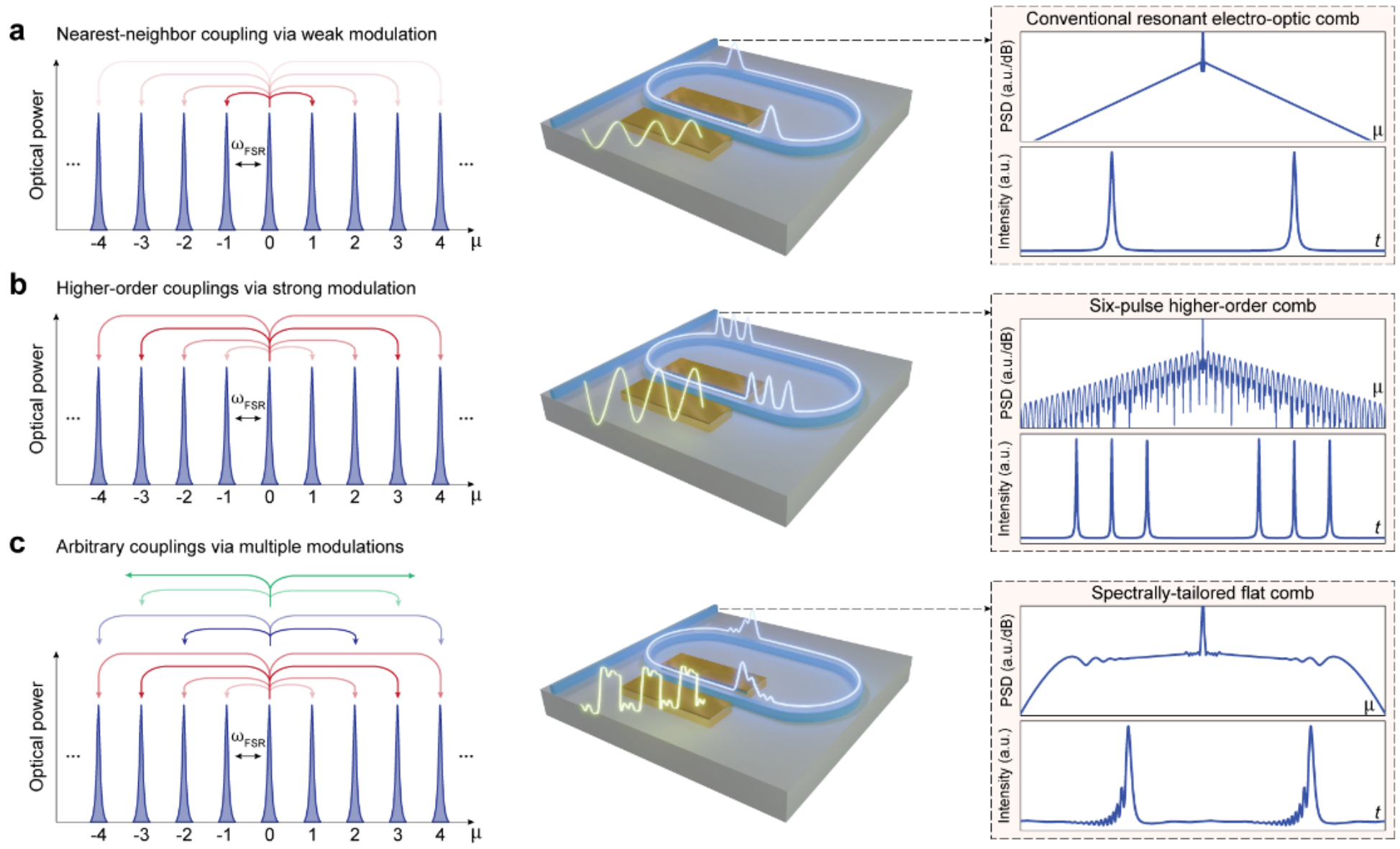

**Fig. 1 | Resonant electro-optic frequency combs on thin-film lithium niobate.** **a**, Device operated in the low modulation depth ($\beta_1 = \pi V_1 / V_{\pi,1}$) regime. Single-tone modulation at $\omega_{MW} = 1 \cdot \omega_{FSR}$ leads to comb generation dynamics well-described by nearest-neighbor interactions. Coupling strengths are indicated by the color transparency of the arrows originating from pump mode ($\mu = 0$) pointing to neighboring modes. This results in conventional comb states characterized by double-sided exponential decay and two pulses per cavity round trip. **b**, Device operated in the large $\beta_1$ regime. In contrast to **a**, the strong EO interaction supplied by the single-tone shifts the dominant coupling to, as illustrated, the third-nearest-neighbor within a cavity round trip. The steady-state spectral signature is an unconventional, envelope-modulated spectrum which corresponds to the emergence of additional temporal periodicity at times scales of the cavity round trip time. In this case, six coherent pulses circulate within the cavity. **c**, Device operated in the multi-tone modulation regime, consisting of several frequencies that satisfy $\omega_{MW,n} = n \cdot \omega_{FSR}$ for integer $n$. Complex long-range couplings are formed by the modulations, differentiated by the colors of the arrows. Spectral shaping towards resonantly-enhanced flat-top combs is illustrated here, by deliberately engineering the set of $\beta_n$ and $\phi_n$.

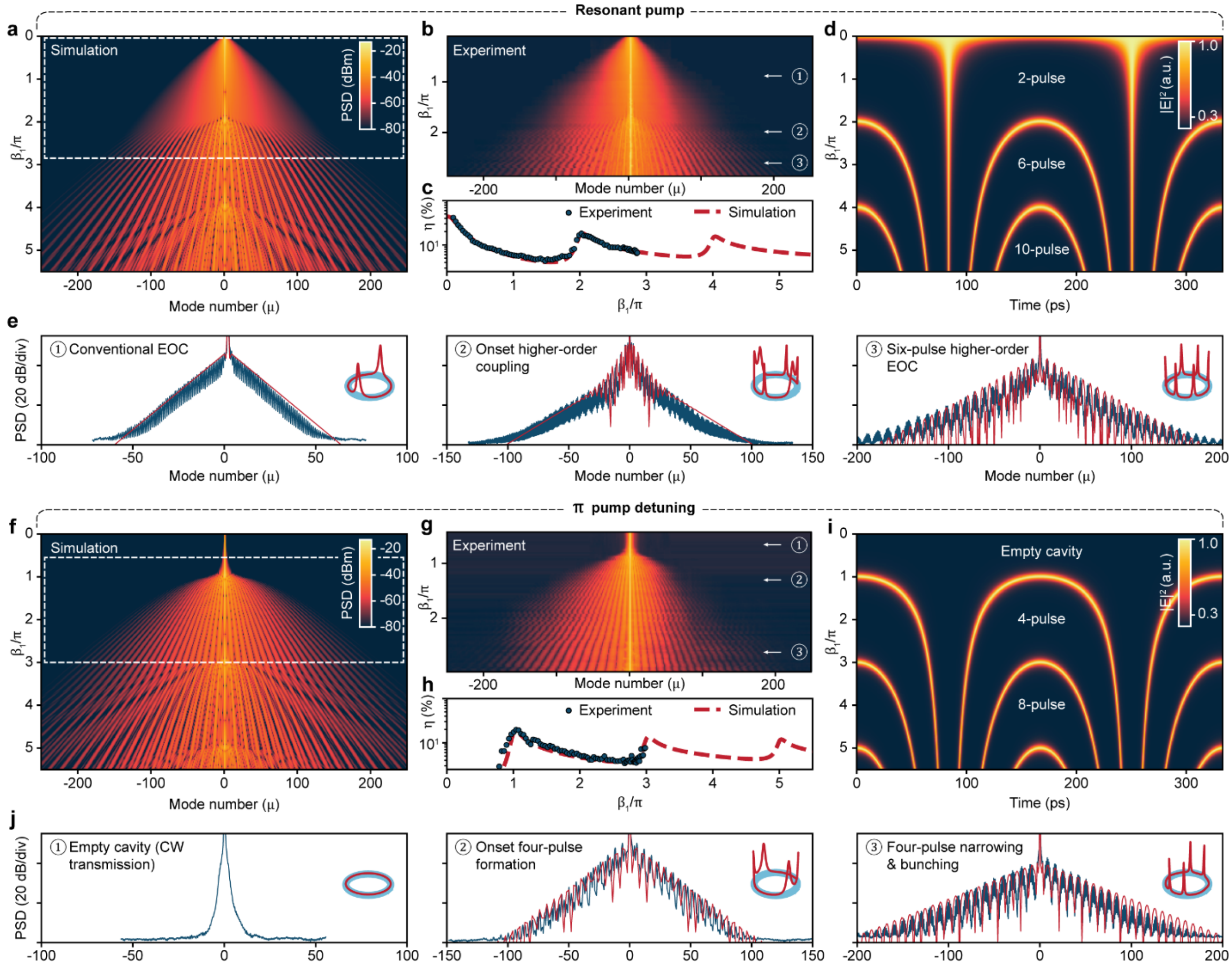

**Fig. 2 | Observation of unconventional electro-optic comb and pulse states. In panels a-e, the optical pump is aligned with a cavity resonance mode, while in f-j it is centered between resonance modes. a**, Simulated spectrogram of resonant EOFCs extracted by the bus waveguide, plotted as a function of $\beta_1$ and $\mu$. The $0 \le \beta_1 < 2\pi$, $2\pi < \beta_1 < 4\pi$, and $\beta_1 > 4\pi$ regimes correspond to distinct nonlinear optical states. The white dashed line indicates the regime accessible in our experiments. **b**, Experimental spectrogram obtained by increasing the microwave voltage (power) applied to the traveling-wave modulator, thus increasing $\beta_1$. The modulation frequency is equal to the fundamental FSR of the cavity, given by $\omega_{MW} = 2\pi \cdot 3.1410$ GHz. The simulation and experiment are in excellent agreement, confirming the validity of our model. Slight deviations in the form of spectral asymmetry are likely due to a perturbation in the cavity dispersion. **c**, Percent pump-to-comb conversion efficiency $\eta$ plotted as a function of $\beta_1$, which features periodic revivals at subsequent $\beta_1$ equal to even multiples of $\pi$, predicted in simulation (red dashed line) and confirmed by experiment (blue dots). The $\eta$ is defined by $\eta = \frac{P_{comb}}{P_p} \cdot 100\%$ where $P_{comb}$ is the total sideband power ($\sum_{\mu \neq 0} P_\mu$) and $P_p = P_0$ is the total pump power. **d,** Simulated pulse patterns corresponding to the spectrogram in **a**. Each horizontal line cut represents the optical field intensity within the cavity, for a specific value of $\beta_1$. Bright regions have high field intensity and correspond to formed pulses, thus confirming that transitions in the spectral envelope are due to the emergence of additional pulse pairs circulating in the cavity. **e**, Select comb spectra at three characteristic $\beta_1$ labeled in **b**. Red lines are simulated envelopes and blue lines are experimental data. Left panel shows a conventional resonant EOFC, middle panel shows the onset of two- to -six-pulse transition, and right shows a six-pulse state. Insets show the simulated intracavity pulse pattern with periodic boundaries applied. **f**, Simulated spectrogram when the pump is maximally detuned from $\mu = 0$, distinguished by the $0 \le \beta_1 < \pi$ (no comb state), $\pi < \beta_1 < 3\pi$, and $3 \le \beta_1 < 5\pi$ regimes. **g**, Experimental spectrogram, where each slice is collected only after confirming $\Delta_p$ via pump detuning spectroscopy (see Supplementary information for details). **h**, $\eta$ plotted as a function of $\beta_1$, rising above zero around $\beta_1 = 1$ and undergoes periodic revivals at subsequent $\beta_1$ equal to odd multiples of $\pi$.

**i**, Simulated pulse patterns corresponding to the spectrogram in **f**. Again, the emergence of spectra from an empty cavity and subsequent transitions in spectral envelope are explained by pulse pair formation and spatial reorganization. **j**, Select comb spectra at three characteristic $\beta_1$ labeled in **g**. Left panel shows no comb generation, middle panel shows the onset of four-pulse formation, and right shows a four-pulse state. All simulation parameters are derived from device parameters and experimental conditions: intrinsic (extrinsic) loss rates $\kappa_{i(e)} \sim 2\pi \cdot 94$ MHz, microwave detuning $\Delta_{MW} = \omega_{MW} - \omega_{FSR} = 0$, negligible integrated dispersion over the relevant comb bandwidths, and pump power normalized to that of the measurement. The Supplementary information provides all details for simulations and the theoretical models used.

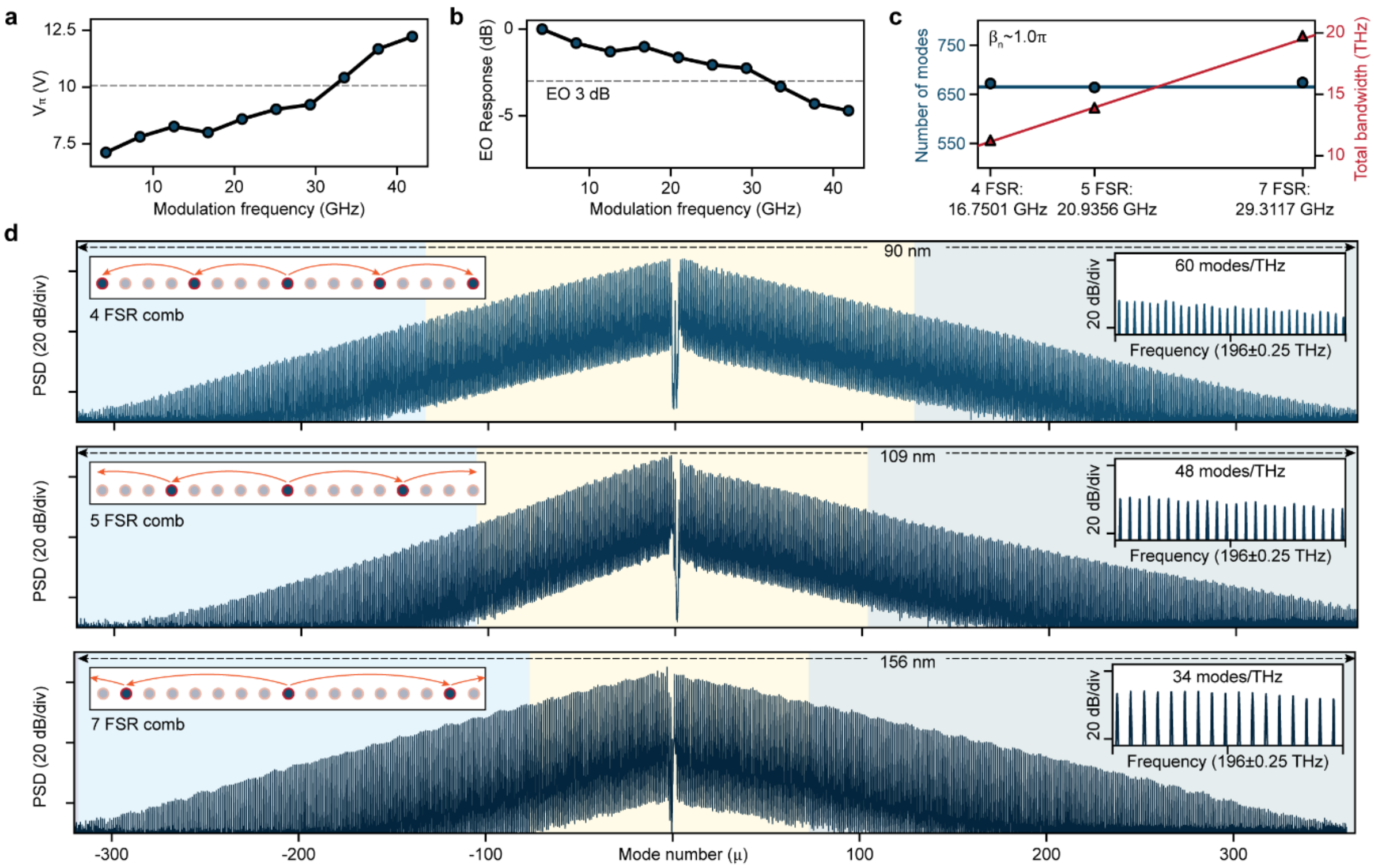

**Fig. 3 | Repetition-rate flexible and broadband electro-optic frequency combs.** **a**, Experimental $V_{\pi,n}$ of the traveling-wave modulated cavity EOFC plotted as a function of modulation frequency $\omega_{MW,n} = n \cdot \omega_{FSR}$ and sampled up to $n = 10$. **b**, Modulation frequency-dependent EO response (i.e., modulation efficiency, or EO-$S_{21}$) where the overall response curve is normalized to the response at $n = 1$ (0 dB level). The device used here is 1.45 cm-long and has a fundamental mode spacing given by $\omega_{FSR} = 2\pi \cdot 4.187$ GHz. Gray dashed lines in **a** and **b** mark $\sqrt{2}V_{\pi,1}$ and -3 dB levels, respectively. **c**, Comb mode and bandwidth scaling plotted as a function of modulation frequency. For a single device, the total mode number (665, blue) is conserved and the total bandwidth scales linearly with modulation frequency (red), provided constant $\beta_n \sim \pi$, for $n = 4$, 5, and 7 shown here. **d**, Multi-FSR-spaced comb spectra corresponding to the three points in **c**. Insets zoom in on half-THz windows about the 196 THz center frequency. Background blue, yellow, and green colors correspond to the L, C, and S bands, respectively. We note that the slight asymmetry in all comb spectra is due to the frequency-dependent dissipation rate of the cavity, featuring higher dissipation, hence steeper slope, at lower mode numbers (i.e., higher wavelengths, or lower frequencies).

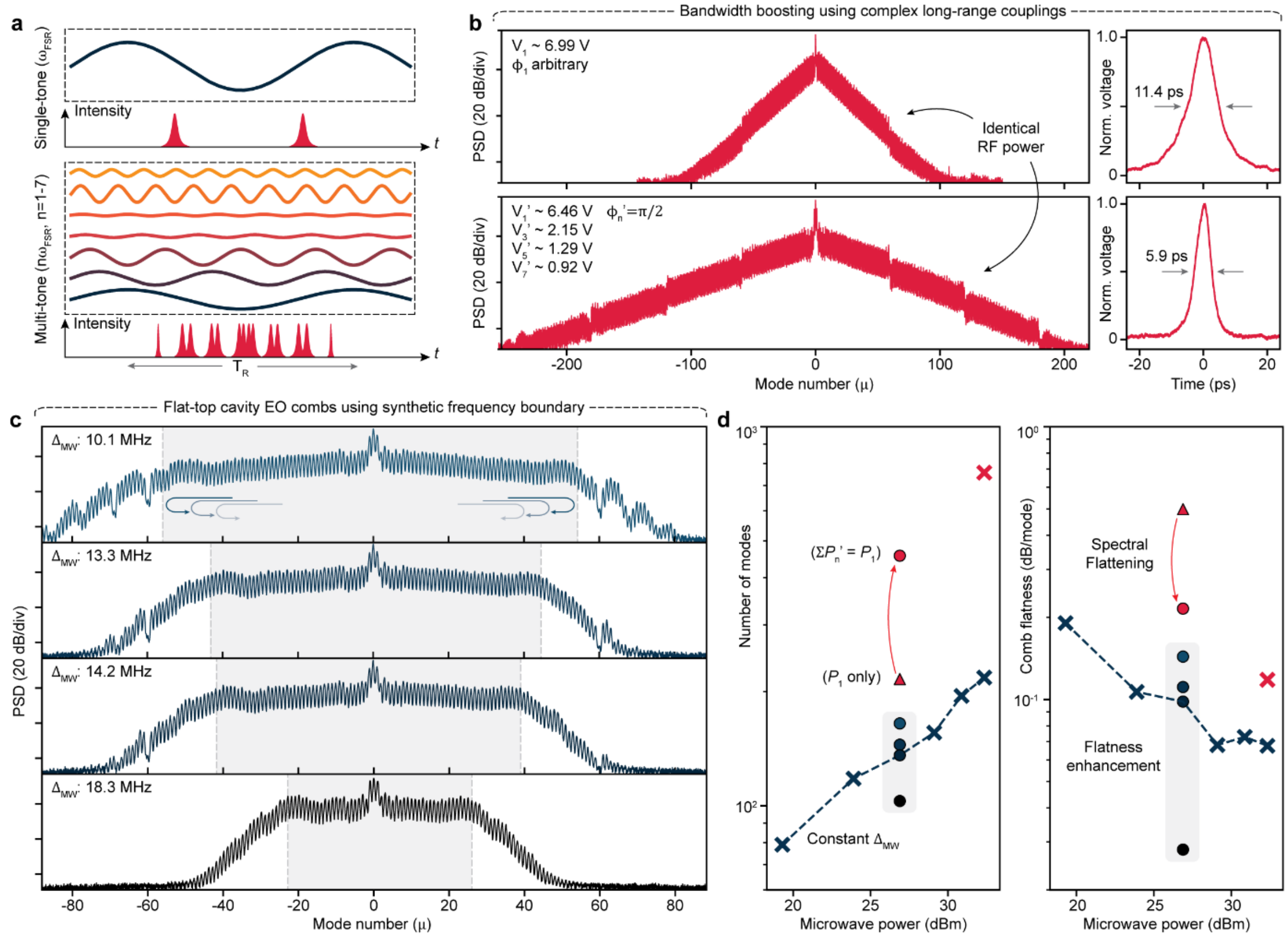

**Fig. 4 | Comb bandwidth boosting and resonantly-enhanced flat-top combs using multi-tone modulation. a,** General illustration of single-tone modulation with modulation frequency given by $\omega_{MW} = 1 \cdot \omega_{FSR} = 2\pi \cdot 4.210$ GHz and multi-tone modulation consisting of seven harmonics of $\omega_{FSR}$ (dashed boxes). The tones can be engineered in both relative amplitude and phase, such that resonant EOFC spectro-temporal properties are actively configurable. **b**, Experimental demonstration of spectral flattening achieved through multi-tone modulation. The top (bottom) panel displays the comb spectrum and autocorrelation trace for single-tone (multi-tone) modulation, resulting in 216 (457) modes and a measured pulse width of 5.7 ps (2.9 ps). We note that the pulses are not compensated for dispersion in the collection path. Pulse width is calculated as half the autocorrelation full-width-at-half-maximum, using a deconvolution factor of 2 for Lorentzian pulses. The total microwave power used is 26.89 dBm in both cases, calculated from the voltages applied as labeled. **c**, Resonantly-enhanced flat-top combs achieved through synthetic frequency boundaries realized by microwave detunings ($\Delta_{MW}$) between 10.1 and 18.3 MHz, in addition to multi-tone modulation as in the bottom panel of **b**. **d**, Number of comb modes generated (left) and comb flatness (right) plotted as a function of total microwave power. Red markers indicate $\Delta_{MW} = 0$, while other markers correspond to $\Delta_{MW} \neq 0$. Red triangle (circle) represents data from the top (bottom) panel of **b**. Dark blue circles within the gray-shaded areas correspond to the four spectra shown in **c**. Dark blue crosses are operated at $\Delta_{MW} = 14.2$ MHz and different microwave power levels than 26.89 dBm. Their spectra, along with that of the red cross, are shown in the Supplementary information.